\documentclass[5p,,preprint,12pt]{elsarticle}
\makeatletter\if@twocolumn\PassOptionsToPackage{switch}{lineno}\else\fi\makeatother

\usepackage{tabulary,xcolor}
\usepackage{amsfonts,amsmath,amssymb}
\usepackage[T1]{fontenc}
\makeatletter
\let\save@ps@pprintTitle\ps@pprintTitle
\def\ps@pprintTitle{\save@ps@pprintTitle\gdef\@oddfoot{\footnotesize\itshape \null\hfill\today}}
\def\hlinewd#1{%
  \noalign{\ifnum0=`}\fi\hrule \@height #1%
  \futurelet\reserved@a\@xhline}
\def\tbltoprule{\hlinewd{.8pt}\\[-12pt]}
\def\tblbottomrule{\noalign{\vspace*{6pt}}\hline\noalign{\vspace*{2pt}}}
\def\tblmidrule{\noalign{\vspace*{6pt}}\hline\noalign{\vspace*{2pt}}}
\AtBeginDocument{\ifNAT@numbers \biboptions{sort&compress}\fi}
\makeatother

\usepackage{ifluatex}
\ifluatex
\usepackage{fontspec}
\defaultfontfeatures{Ligatures=TeX}
\usepackage[]{unicode-math}
\unimathsetup{math-style=TeX}
\else 
\usepackage[utf8]{inputenc}
\fi 
\ifluatex\else\usepackage{stmaryrd}\fi

\usepackage{url,multirow,morefloats,floatflt,cancel,tfrupee}
\makeatletter

\AtBeginDocument{\@ifpackageloaded{textcomp}{}{\usepackage{textcomp}}}
\makeatother
\usepackage{colortbl}
\usepackage{xcolor}
\usepackage{pifont}
\usepackage[nointegrals]{wasysym}
\makeatletter

\def\mcWidth#1{\csname TY@F#1\endcsname+\tabcolsep}

\def\cAlignHack{\rightskip\@flushglue\leftskip\@flushglue\parindent\z@\parfillskip\z@skip}
\def\rAlignHack{\rightskip\z@skip\leftskip\@flushglue \parindent\z@\parfillskip\z@skip}

\@ifundefined{etal}{}{}

\usepackage{ifxetex}
\ifxetex\else\if@twocolumn\@ifpackageloaded{stfloats}{}{\usepackage{dblfloatfix}}\fi\fi

\AtBeginDocument{
\expandafter\ifx\csname eqalign\endcsname\relax
\def\eqalign#1{\null\vcenter{\def\\{\cr}\openup\jot\m@th
  \ialign{\strut$\displaystyle{##}$\hfil&$\displaystyle{{}##}$\hfil
      \crcr#1\crcr}}\,}
\fi
}

\AtBeginDocument{%
  \@ifpackageloaded{endfloat}%
   {\renewcommand\efloat@iwrite[1]{\immediate\expandafter\protected@write\csname efloat@post#1\endcsname{}}}{\newif\ifefloat@tables}%
}%

\def\BreakURLText#1{\@tfor\brk@tempa:=#1\do{\brk@tempa\hskip0pt}}
\let\lt=<
\let\gt=>
\def\processVert{\ifmmode|\else\textbar\fi}

\@ifundefined{subparagraph}{
\def\subparagraph{\@startsection{paragraph}{5}{2\parindent}{0ex plus 0.1ex minus 0.1ex}%
{0ex}{\normalfont\small\itshape}}%
}{}

\newcommand\role[1]{\unskip}
\newcommand\aucollab[1]{\unskip}
  
\@ifundefined{tsGraphicsScaleX}{\gdef\tsGraphicsScaleX{1}}{}
\@ifundefined{tsGraphicsScaleY}{\gdef\tsGraphicsScaleY{.9}}{}
\def\checkGraphicsWidth{\ifdim\Gin@nat@width>\linewidth
	\tsGraphicsScaleX\linewidth\else\Gin@nat@width\fi}

\def\checkGraphicsHeight{\ifdim\Gin@nat@height>.9\textheight
	\tsGraphicsScaleY\textheight\else\Gin@nat@height\fi}

\def\fixFloatSize#1{}%\@ifundefined{processdelayedfloats}{\setbox0=\hbox{\includegraphics{#1}}\ifnum\wd0<\columnwidth\relax\renewenvironment{figure*}{\begin{figure}}{\end{figure}}\fi}{}}
\let\ts@includegraphics\includegraphics

\def\inlinegraphic[#1]#2{{\edef\@tempa{#1}\edef\baseline@shift{\ifx\@tempa\@empty0\else#1\fi}\edef\tempZ{\the\numexpr(\numexpr(\baseline@shift*\f@size/100))}\protect\raisebox{\tempZ pt}{\ts@includegraphics{#2}}}}

\AtBeginDocument{\def\includegraphics{\@ifnextchar[{\ts@includegraphics}{\ts@includegraphics[width=\checkGraphicsWidth,height=\checkGraphicsHeight,keepaspectratio]}}}

\DeclareMathAlphabet{\mathpzc}{OT1}{pzc}{m}{it}

\def\URL#1#2{\@ifundefined{href}{#2}{\href{#1}{#2}}}

\def\UrlOrds{\do\*\do\-\do\~\do\'\do\"\do\-}%
\g@addto@macro{\UrlBreaks}{\UrlOrds}

\edef\fntEncoding{\f@encoding}

\makeatother

\newif\ifmultipleabstract\multipleabstractfalse%
\usepackage[flushleft]{threeparttablex}

\usepackage{float}

\newcommand{\texttildeapprox}{{\fontfamily{pcr}\selectfont\texttildelow}}

\begin{document}

\begin{frontmatter}

\title{
    EnvGAN: Adversarial Synthesis of Environmental Sounds for Data Augmentation    
}
    
\author[abd29c1e65e94]{Aswathy Madhu\corref{c-6ea4d82102b2}}
\ead{aswathymadhu@cet.ac.in}\cortext[c-6ea4d82102b2]{Corresponding author.}
\author[a67edc4a90329]{Suresh K}
\ead{sureshk@cet.ac.in}
    
\address[abd29c1e65e94]{
    College of Engineering, Trivandrum}
  	
\address[a67edc4a90329]{
    Govt. Engineering College, Barton Hill, Thiruvananthapuram}

\begin{abstract}
 The research in Environmental Sound Classification (ESC) has been progressively growing with the emergence of deep learning algorithms. However, data scarcity poses a major hurdle for any huge advance in this domain. Data augmentation offers an excellent solution to this problem. While Generative Adversarial Networks (GANs) have been successful in generating synthetic speech and sounds of musical instruments, they have hardly been applied to the generation of environmental sounds. This paper presents EnvGAN, the first ever application of GANs for the adversarial generation of environmental sounds. Our experiments on three standard ESC datasets illustrate that the EnvGAN can synthesize audio similar to the ones in the datasets. The suggested method of augmentation outshines most of the futuristic techniques for audio augmentation.
\end{abstract}

\end{frontmatter}
    
\section{Introduction}
 Environmental Sound Classification (ESC) has been a dynamic field of research in the domain of audio signal processing for the last few years. By environmental sound, we mean any audible sound event which is irrelevant to the listener. They are listened mainly for source identification. Automatic ESC plays a pivotal role in many applications like query by audio content\unskip~\cite{864172:20206024}, automatic identification of tags for audio\unskip~\cite{864172:20206029}, home automation\unskip~\cite{864172:20206008}, remote surveillance\unskip~\cite{864172:20206023} etc. A plethora of algorithms are available for the automatic classification of environmental sounds. In particular, deep learning algorithms are persistently acquiring momentum in this field. Yet the paucity of publicly accessible datasets retards each and every rapid stride in the domain.

 Data augmentation offers an excellent solution to this issue. Data augmentation means training the deep network with additional diverse data. This increases the generalization capability of the network and reduces overfitting. The traditional method of data augmentation utilizes elementary transformations applied in such a way that these transformations do not alter the semantic essence of the associated target labels. For instance, even after time shifting, a siren would still be a siren. By training the network in this manner, its response towards the real-time data is improved.

Jaitley and Hinton\unskip~\cite{864172:20206028} introduced the idea of augmentation in the domain of speech on account of its positive results in object recognition\unskip~\cite{864172:20206001}. Other successful speech augmentation methods include \unskip~\cite{864172:20206019,864172:20205990,864172:20206262}. Similar attempts were made in music processing also\unskip~\cite{864172:20205987,864172:20206004,864172:20206264}. Regardless of the motivating results of data augmentation in these fields, it has limited applications in environmental sound classification\unskip~\cite{864172:20206010,864172:20205997}.

The major downside of traditional augmentation techniques is that there can be a variety of real world conditions which cannot be justified by these simple techniques. This is the ground for using GANs in data augmentation. An added advantage of GAN is that it can efficiently handle class imbalanced datasets\unskip~\cite{864172:21562743} . Some remarkable implementations of image augmentation using GANs are \unskip~\cite{864172:20206022,864172:20206017,864172:20234028,864172:20206013}. In spite of the positive results in image processing applications, audio data augmentation using GANs is in its emergence. Donahue et.al.\unskip~\cite{864172:20205988} made the first attempt of using GAN for the synthetic audio generation. They synthesized audio from a variety of domains like speech, bird sounds, and instrumental sounds. Lee et.al. \unskip~\cite{864172:20206007} explored the use of conditional GANs for synthetic audio generation. 

The contributions of this work is threefold. (1) We present a high capacity deep CNN model for environmental sound classification. Our model is derived from \unskip~\cite{864172:20206005} with some significant changes as described in section 4.2. We demonstrate that our baseline model outperforms most of the previous models including the one described in\unskip~\cite{864172:20206005}. We observe that the performance of the model can further be improved by data augmentation. This is validated by the experimental results of the model trained with traditional data augmentation. (2) We investigate a time domain strategy for synthetic generation of environmental sounds using GAN (EnvGAN) to account for the real world conditions that cannot be justified by traditional augmentation. We notice that the EnvGAN can generate a wide variety of environmental sounds similar to the ones in the dataset. (3) We evaluate the utility of EnvGAN as an augmentation tool in the context of automatic ESC using deep CNN. The proposed classification model with augmentation is assessed on three standard ESC datasets - ESC-10, UrbanSound8K and TUT Urban Acoustic Scenes 2018 development dataset to gauge its performance. We prove that the recommended method outstrips most of the sophisticated audio augmentation techniques for automatic environmental sound classification. A preliminary version of this work has appeared in \unskip~\cite{864172:20205999}, and this is an extended version with comprehensive experimental evaluations. 

The remainder of the paper is organized as follows. In section 2, we present some preliminaries of GAN along with the model architecture and parameters of the proposed EnvGAN.  Section 3 describes the datasets employed in this work. Section 4 outlines the proposed high capacity deep CNN model for ESC. Section 5 delineates the experimental outcomes and consequent discussions. In Section 6, we conclude the paper with interesting avenues for future research.
    
\section{Proposed GAN model for Augmentation}

\subsection{GAN Preliminaries}GAN is a deep neural network architecture consisting of two networks competing against each other. Ian Goodfellow\unskip~\cite{864172:20206018} introduced these networks in the literature in 2014. In his formulation, a generator network which generates new data instances is pitted against a discriminator which evaluates the data instances for authenticity. Training these networks corresponds to the well known minimax strategy for training a zero-sum game where
\let\saveeqnno\theequation
\let\savefrac\frac
\def\dispfrac{\displaystyle\savefrac}
\begin{eqnarray}
\let\frac\dispfrac
\gdef\theequation{1}
\let\theHequation\theequation
\label{dfg-37f15f61b193}
\begin{array}{@{}l}V(G,D)=E_{x \sim P_{data}(x)}[log (D(x))]+\\ \qquad  \qquad E_{z \sim P_{z}(z)}[log(1-D(G(z)))]
\end{array}
\end{eqnarray}
\global\let\theequation\saveeqnno
\addtocounter{equation}{-1}\ignorespaces 
is the value function. G tries to minimize V(G, D) whereas D tries to maximize it. At perfect equilibrium, the generator would capture the general training data distribution and D would be 1/2 everywhere. In other words, the discriminator would be always unsure of whether its inputs are real or fake. Goodfellow demonstrates that their training algorithm actually minimizes the information radius (Jensen Shannon divergence) between the distributions of data and generator. However, if the generating process is far away from the ground truth, the generator gradients will vanish and it will not learn anything. 

To tackle the vanishing gradient problem, Arjovsky et. al.\unskip~\cite{864172:20205996}  proposed a new cost function using Wasserstein distance (WGAN) that has a smoother gradient everywhere. To calculate the Wasserstein distance, a 1-Lipschitz function is needed. They suggested weight clipping as a method of enforcing this constraint on the discriminator model. But the model performance was very sensitive to the clipping parameter. Hence, Gulrajani et.al.\unskip~\cite{864172:20206003} suggested gradient penalty (WGAN-GP) as an alternative strategy to enforce Lipschitz constraint. They demonstrate that WGAN-GP enhances training stability and produces good results even when other GAN losses fail.

\subsection{EnvGAN}The EnvGAN architecture is based on WaveGAN\unskip~\cite{864172:20205988} which introduced GANs for audio synthesis. In the original WaveGAN formulation, the model generates 16384 samples (just above 1 sec at a sampling rate of 16000 Hz). This duration is sufficient for some fields (for example, special effects) while this length is not sufficient for some other domains. For ESC, human listeners could correctly identify the environmental sounds with an acceptable accuracy of 82\% when the audio clip was 4-sec long (based on a listening test conducted in \unskip~\cite{864172:20205991}). Hence, to generate 4-sec audios at 44.1 kHz, we add two layers to WaveGAN resulting in an output length of 196608 samples. Our model generates monaural audio.

\subsubsection{Model Architecture}A complete description of the EnvGAN architecture is given below:

 The input to the generator is a random sample taken from a uniform distribution between -1 and 1. The input layer reshapes the random sample into 16\ensuremath{\times}12288. This is followed by seven up-convolution layers which gradually converts a course input feature map to a fine and detailed output. In Table~\ref{tw-e6fd6a87e53b}, we list the full architecture for the EnvGAN generator.

\begin{table*}[!htbp]
\caption{{EnvGAN generator architecture} }
\label{tw-e6fd6a87e53b}
\def\arraystretch{1}
\ignorespaces 
\centering 
\begin{tabulary}{\linewidth}{p{\dimexpr.3368\linewidth-2\tabcolsep}p{\dimexpr.30059999999999995\linewidth-2\tabcolsep}p{\dimexpr.36260000000000005\linewidth-2\tabcolsep}}
\tbltoprule Operation & Kernel size & Output Shape\\
\tblmidrule 
Input \mbox{}\protect\newline  z{\texttildeapprox} Uniform(-1,1) &
   &
  (b,100)\\
Dense\_1 &
  (100, 3072d) &
  (b, 3072d)\\
Reshape\_1 &
   &
  (b, 16, 192d)\\
ReLU\_1 &
   &
  (b, 16, 192d)\\
Up\_Conv\_1 (s = 4) &
  (25, 192d, 96d) &
  (b, 64, 96d)\\
ReLU\_2 &
   &
  (b, 64, 96d)\\
Up\_Conv\_2 (s = 4) &
  (25, 96d, 48d) &
  (b, 256, 48d)\\
ReLU\_3 &
   &
  (b, 256, 48d)\\
Up\_Conv\_3 (s = 4) &
  (25, 48d, 24d) &
  (b, 1024, 24d)\\
ReLU\_4 &
   &
  (b, 1024, 24d)\\
Up\_Conv\_4 (s = 4) &
  (25, 24d, 12d) &
  (b, 4096, 12d)\\
ReLU\_5 &
   &
  (b, 4096, 12d)\\
Up\_Conv\_5 (s = 4) &
  (25, 12d, 6d) &
  (b, 16384, 6d)\\
ReLU\_6 &
   &
  (b, 16384, 6d)\\
Up\_Conv\_6 (s = 4) &
  (25, 6d, 3d) &
  (b, 65536, 3d)\\
ReLU\_7 &
   &
  (b, 65536, 3d)\\
Up\_Conv\_7 (s = 3) &
  (25, 3d, c) &
  (b, 196608, c)\\
Tanh &
   &
  (b, 196608, c)\\
\tblbottomrule 
\end{tabulary}\par 
\end{table*}
The transposed convolution (up-convolution) operation in the generator upsamples the input feature map (by inserting zeros) without losing many details (by applying a learned filter bank). During optimization, the learned filters are adapted such that they reflect the underlying relationship within the dataset. However, transposed convolution produces checkerboard artifacts\unskip~\cite{864172:20205988}. To ensure that the discriminator does not learn these artifacts, we use phase shuffle operation (with hyper parameter n=2) as suggested in \unskip~\cite{864172:20205988}. 

The discriminator is a CNN which checks the generated samples for authenticity. Its input is an audio sample of dimension196608\ensuremath{\times}1. Table~\ref{tw-12a3afc6467e} lists the full architecture of the discriminator. In these tables, 'b', 'd', and 'c' denote batch size, dimension of EnvGAN and the number of input channels respectively. 
\begin{table*}[!htbp]
\caption{{EnvGAN discriminator architecture} }
\label{tw-12a3afc6467e}
\def\arraystretch{1}
\ignorespaces 
\centering 
\begin{tabulary}{\linewidth}{p{\dimexpr.4423\linewidth-2\tabcolsep}p{\dimexpr.2536\linewidth-2\tabcolsep}p{\dimexpr.30410000000000004\linewidth-2\tabcolsep}}
\tbltoprule Operation & Kernel Size & Output Shape\\
\tblmidrule 
Input &
   &
  (b, 196608, c)\\
Conv\_1 (s = 3) &
  (25, c, 3d) &
  (b, 65536, 3d)\\
Leaky ReLU\_1 $(\alpha=0.2) $ &
   &
  (b, 65536, 3d)\\
Phase Shuffle\_1 &
   &
  (b, 65536, 3d)\\
Conv\_2 (s = 4) &
  (25, 3d, 6d) &
  (b, 16384,  6d)\\
Leaky ReLU\_2 $(\alpha=0.2) $ &
   &
  (b, 16384,  6d)\\
Phase Shuffle\_2  &
   &
  (b, 16384,  6d)\\
Conv\_3 (s = 4) &
  (25, 6d, 12d) &
  (b, 4096, 12d)\\
Leaky ReLU\_3 $(\alpha=0.2) $ &
   &
  (b, 4096, 12d)\\
Phase Shuffle\_3  &
   &
  (b, 4096, 12d)\\
Conv\_4 (s = 4) &
  (25, 12d, 24d) &
  (b, 1024, 24d)\\
Leaky ReLU\_4 $(\alpha=0.2) $ &
   &
  (b, 1024, 24d)\\
Phase Shuffle\_4 &
   &
  (b, 1024, 24d)\\
Conv\_5 (s = 4) &
  (25, 24d, 48d) &
  (b, 256, 48d)\\
Leaky ReLU\_5 $(\alpha=0.2) $ &
   &
  (b, 256, 48d)\\
Phase Shuffle\_5 &
   &
  (b, 256, 48d)\\
Conv\_6 (s = 4) &
  (25, 48d, 96d) &
  (b, 64, 96d)\\
Leaky ReLU\_6 $(\alpha=0.2) $ &
   &
  (b, 64, 96d)\\
Phase Shuffle\_6 &
   &
  (b, 64, 96d)\\
Conv\_7 (s = 4) &
  (25, 96d, 192d) &
  (b, 16, 192d)\\
Leaky ReLU\_7 $(\alpha=0.2) $ &
   &
  (b, 16, 192d)\\
Phase Shuffle\_7 &
   &
  (b, 16, 192d)\\
Reshape\_1 &
   &
  (b, 3072d)\\
Dense\_1 &
  (3072d, 1) &
  (b, 1)\\
\tblbottomrule 
\end{tabulary}\par 
\end{table*}

\subsubsection{Model Parameters} The values of parameters and hyperparameters associated with EnvGAN for our experiments are listed in Table~\ref{tw-c414ee7cf710}.
\begin{table}[!htbp]
\caption{{EnvGAN parameters and hyperparameters} }
\label{tw-c414ee7cf710}
\def\arraystretch{1}
\ignorespaces 
\centering 
\begin{tabulary}{\linewidth}{p{\dimexpr.4461\linewidth-2\tabcolsep}p{\dimexpr.5539\linewidth-2\tabcolsep}}
\tbltoprule Name & Value\\
\tblmidrule 
Number of channels &
  1\\
Sampling rate &
  44100\\
Generation length &
  196608\\
Batch size &
  64\\
Kernel length &
  25\\
Latent dimension &
  100\\
D updates per G update &
  5\\
Optimizer &
  Adam $(\alpha=10^{-4},\beta_1=0.5,\beta_2=0.9) $\\
Loss &
  WGAN-GP $(\lambda=10) $\\
\tblbottomrule 
\end{tabulary}\par 
\end{table}

\section{Dataset}
Three benchmark datasets were used to evaluate the performance of the proposed method.  A brief description of each dataset is given below. The evaluation strategy in three datasets is 10-fold cross validation.

\subsection{ESC-10}ESC-10\unskip~\cite{864172:20206021} is a collection of 400 labeled environmental recordings equally balanced between 10 classes (40 clips per class). These classes are \textit{dog barking (DO), rainfall (RA), waves of sea (SE), crying baby (BA), clock ticking(CL), sneezing (PE), helicopter (HE), chainsaw (CH), crowing rooster (RO) and fire crackling (FI)}. The dataset consists of 5-second sound clips created from recordings in the Freesound project\unskip~\cite{864172:20206026}.The inter class differences are so striking that there is only limited ambiguity in classification

\subsection{UrbanSound8K}UrbanSound8K \unskip~\cite{864172:20206027} is an imbalanced collection of 8732 clips (\ensuremath{\leq } 4s) of sounds from cities pertaining to ten low level classes derived from urban sound taxonomy. These classes are \textit{air conditioner (AI), car horn (CA), children playing (CH), dog barking (DO), drilling (DR), engine idling (EN), gunshot (GU), jackhammer (JA), siren (SI), and street music (ST)}. The audio slices (in .wav format) are prearranged into 10 folds. The corresponding metadata file is available in .csv format.

\subsection{TUT Urban Acoustic Scenes 2018 development dataset}TUT Urban Acoustic Scenes 2018 development dataset\unskip~\cite{864172:20205994} consists of 8640 10-seconds audio segments equally balanced between 10 classes (864 segments each): \textit{airport (AP), }\textit{shopping mall (SH), }\textit{metro station (MS), }\textit{street pedestrian (SP), }\textit{public square (PU), }\textit{street traffic (ST), }\textit{tram (TR),}\textit{bus (BU), }\textit{metro (ME) and } \textit{park (PA).}

The details of these datasets including the number of labels and the number of audio clips are summarized inTable~\ref{tw-a5419b669eaf}. In this work, the datasets are prearranged to 10 folds for the ease of 10-fold cross validation.
\begin{table}[!htbp]
\caption{{Summary of datasets} }
\label{tw-a5419b669eaf}
\def\arraystretch{1}
\ignorespaces 
\centering 
\begin{tabulary}{\linewidth}{p{\dimexpr.3149\linewidth-2\tabcolsep}p{\dimexpr.2025\linewidth-2\tabcolsep}p{\dimexpr.1853\linewidth-2\tabcolsep}p{\dimexpr.2973\linewidth-2\tabcolsep}}
\tbltoprule Datasets & \# classes & \# clips & Experimental setup\\
\tblmidrule 
ESC-10 &
  10 &
  400 &
  10-fold cross validation\\
UrbanSound8K &
  10 &
  8732 &
  10-fold cross validation\\
TUT &
  10 &
  8640 &
  10-fold cross validation\\
\tblbottomrule 
\end{tabulary}\par 
\end{table}

\section{Proposed Environmental Sound Classification Model}

\subsection{Data Preprocessing}Raw audio signal is not an appropriate input to a classifier, even if it is a Deep Neural Network due to two reasons (1) Audio signals are extremely high dimensional (2) Perceptually similar sounds may not be neighbors in vector space\unskip~\cite{864172:20206011}. Hence, even for systems that employ feature learning the audio signal should be transformed into an appropriate representation that ensures successful learning. We adopted the most popular method for signal representation (log mel spectrogram). We use librosa - a python library which provides basic routines for music information retrieval and audio analysis\unskip~\cite{864172:20205993}-to read and process the raw audio signal. We use log scaled mel spectrogram to transform the raw audio signal to the attribute space. The log scaled mel spectrogram is computed with128 mel frequency bands as it is a reasonable size that sufficiently retains the original spectral characteristics while significantly reducing the dimensionality of the data. We use a frame length of 23ms (1024 samples at 44100 Hz) and equal hop size. To cope with the varying sampling rate of samples in the dataset, all recordings are resampled to 44100 Hz (audio CD's sampling rate). Generally 20 kHz is the upper threshold of human hearing range. This makes 44100 Hz a sensible choice for sampling rate. Handcrafted feature extraction is performed on a frame level with a frame length of 23ms. Non overlapping frames are used since we are not concerned about faithfully recreating the audio, but about the feature values only.  128 frames are selected from (initial frame is selected randomly) the overall log scaled mel spectrogram resulting in 128\ensuremath{\times}128\ensuremath{\times}1 input to the classifier.

\subsection{Deep CNN classifier}The deep CNN employed here is a modified form of \unskip~\cite{864172:20206005} with the following significant changes. (1) Optimizer was changed from SGD to Adam with learning rate 0.001 (2) Dropout was removed from the output layer and added to the flatten layer before the last hidden layer (in which the dropout was retained) (3) After a series of experiments, a dropout rate of 0.5 was adopted in the last hidden layer and the flatten layer with maxnorm constraint on the weights of the last hidden layer. (4) The minibatch size was changed from 100 to 128. (5) The number of epochs was changed from 50 to 150. During training, the network learns the model parameters which helps in mapping the input audio to the associated label.

The CNN architecture has 3 convolutional layers employing 24, 48 and 48 two-dimensional filters of size 5\ensuremath{\times}5. The first two convolutional layers are stacked with pooling layers. We use max-pooling (s = (4,2)) and ReLU activation for the convolutional layers. This is followed by reshape/flatten and dense layer with 64 units and ReLU activation. Dropout is introduced in reshape layer and dense layer with a rate of 0.5 to prevent the network from overfitting. The output layer is a dense layer (10 units) employing softmax activation to return the probabilities of each class. The summary of the model along with the parameter count in each layer is shown in Figure~\ref{f-f396a84f9606}.

\bgroup
\fixFloatSize{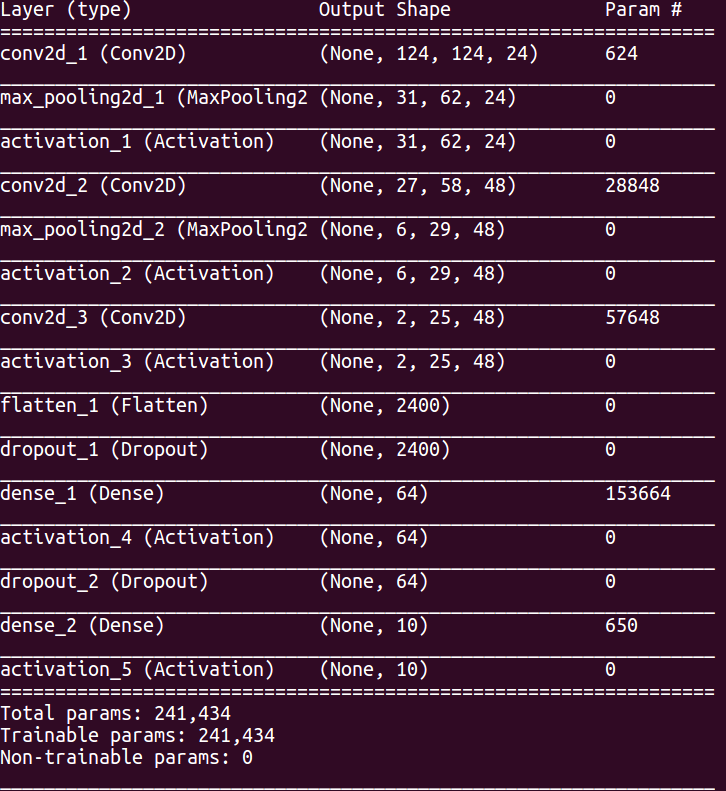}
\begin{figure}[!htbp]
\centering \makeatletter\IfFileExists{images/cnn_architecture.png}{\includegraphics[width=.92\linewidth]{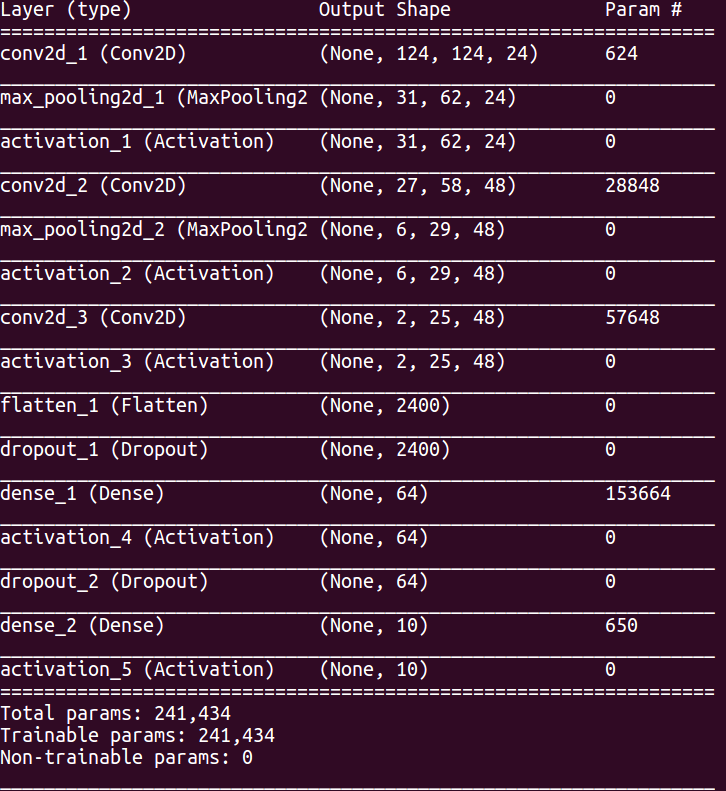}}{}
\makeatother 
\caption{{CNN Architecture} \mbox{}\protect\newline}
\label{f-f396a84f9606}
\end{figure}
\egroup
For training, the loss function used is categorical cross-entropy which enhances the ease of interpretation. Categorical cross-entropy is an indicator of the probability of a sample belonging to the target class. The model is optimized using Adam optimizer (learning rate = 0.001). The training is terminated after 150 epochs.The CNN is implemented in Python with Keras\unskip~\cite{864172:20206016}.

\subsection{Data Augmentation}We implemented four traditional augmentation techniques - time stretching, pitch shifting, dynamic range compression and additive background noise\unskip~\cite{864172:20206005}. This resulted in five different augmentation sets. The augmentations are applied to the audio signal before preprocessing. The augmentation parameters are chosen such that they do not change the identity of the associated label. The employed augmentation techniques are listed below. \mbox{}\protect\newline 1) Time\_stretching (time\_stretch): alters the duration of the audio sample without affecting the pitch. Four stretching factors are used: \{0.85, 0.95, 1.05, 1.15\}. \mbox{}\protect\newline 2) Pitch\_shifting (pitch\_shift\_1, pitch\_shift\_2): alters the pitch of the audio sample without affecting the duration. Eight shifting factors are used: \{-2, -1.5, -1, -0.5, 0.5, 1, 1.5, 2\}. \mbox{}\protect\newline 3) Additive background noise (background): 4 acoustic scenes \{\textit{background\_noise, football\_crowd, elaborate\_thunder, creepy\_background}\} are added to each audio sample. Each mix is generated by the equation z~= x + w.y. Here 'x' is the original audio, 'y' is the background acoustic scene and 'w' is a weighting parameter uniformly distributed between 0.1 and 0.5. Care has been taken to ensure that the background scenes do not contain any of the target classes. \mbox{}\protect\newline 4) Dynamic range compression (drcomp): The process of suppressing loud sounds or enhancing quiet sounds thereby reducing the dynamic range of the input audio is called Dynamic Range Compression. In this paper, dynamic range compression was implemented using SoX (Sound eXchange)\unskip~\cite{864172:20206315} with 4 general profiles \{speech, podcast, music, voice/radio\} all taken from the Doom9's forum featuring information on digital audio manipulation\unskip~\cite{864172:20206009}.

\subsubsection{Data Augmentation using EnvGAN}In this work, for each dataset, we train the EnvGAN on each class individually for 2500 epochs. The generated files are randomly arranged in ten folders to facilitate 10-fold cross validation. To ensure that EnvGAN is trained sufficiently, a similarity score (S) is computed for the generated sound files to assess their similarity with the original dataset defined by  
\begin{eqnarray*}S={\textstyle\sum_{n=1}^{N}}\frac{(a\lbrack n\rbrack-b\lbrack n\rbrack)^{2}}{a\lbrack n\rbrack b\lbrack n\rbrack} \end{eqnarray*}
where A = \{a[n]\} and B = \{b[n]\} are two audio signals. The similarity threshold is fixed at S = 0.1. The generated files with the similarity score less than this threshold are rejected. The details of the waveform generation including the number of files generated and time for generation (when run in Google Colaboratory with NVIDIA Tesla V100-SXM2 GPU) are summed up in Table~\ref{tw-474fa73c9990}.
\begin{table}[!htbp]
\caption{{Summary of the audio generation using EnvGAN} }
\label{tw-474fa73c9990}
\def\arraystretch{1}
\ignorespaces 
\centering 
\begin{tabulary}{\linewidth}{p{\dimexpr.4199\linewidth-2\tabcolsep}p{\dimexpr.2761\linewidth-2\tabcolsep}p{\dimexpr.304\linewidth-2\tabcolsep}}
\tbltoprule Dataset & \# files & Time\\
\tblmidrule 
ESC-10 &
  5000 &
  {\texttildeapprox}37 min\\
UrbanSound8K &
  30000 &
  {\texttildeapprox}191 min\\
TUT &
  30000 &
  {\texttildeapprox}194 min\\
\tblbottomrule 
\end{tabulary}\par 
\end{table}

\section{Results and Discussions}

\subsection{Performance of different augmentation schemes}We use 10-fold cross validation to gauge the performance of the model. In these experiments, the baseline model denotes the deep CNN trained without augmentation. We use six different evaluation metrics to quantify the results of these experiments as discussed below.

  \begin{enumerate}
  \item \relax  Confusion matrix\unskip~\cite{864172:20206020} : A matrix used to evaluate the quality of the classification model. 
  \item \relax Accuracy\unskip~\cite{864172:20206020} : the degree of closeness of the predicted value to the true value. 
  \item \relax Precision\unskip~\cite{864172:20206020} : the degree of closeness of the predicted values to each other. 
  \item \relax Recall\unskip~\cite{864172:20206020} : the percentage of total relevant results correctly classified by the algorithm.
  \item \relax F1-score\unskip~\cite{864172:20206020} : harmonic mean of precision and recall.
  \item \relax Cohen's kappa score\unskip~\cite{864172:20206000} : the degree of superiority of a classification algorithm over a naive classifier.
  \end{enumerate}
   The results of applying different augmentation schemes to each dataset are summarized in Table~\ref{tw-9699bb46e946}. In the table, the symbols Baseline, time\_stretch, pitch\_shift1, pitch\_shift2, drcomp, background, and GAN denote the model trained on original dataset, Time Stretching, Pitch Shifting by integer and real factors, Dynamic Range Compression, Additive Backgorund Noise and the suggested augmentation scheme using EnvGAN respectively. Each column of the table indicates the mean value of the metric along with its standard deviation indicated within parentheses. It is evident from the table that EnvGAN performs better than the baseline model in all datasets. It surpasses other augmentation techniques in ESC-10 while it gives results on a par with other techniques in UrbanSound8K and TUT.

\begin{table*}[!htbp]
\caption{{Comparison of augmentation schemes} }
\label{tw-9699bb46e946}
\centering 
\begin{threeparttable}

\def\arraystretch{1}
\ignorespaces 
\centering 
\begin{tabulary}{\linewidth}{p{\dimexpr.14\linewidth-2\tabcolsep}p{\dimexpr.14\linewidth-2\tabcolsep}p{\dimexpr.14510000000000002\linewidth-2\tabcolsep}p{\dimexpr.1414\linewidth-2\tabcolsep}p{\dimexpr.14369999999999997\linewidth-2\tabcolsep}p{\dimexpr.14780000000000001\linewidth-2\tabcolsep}p{\dimexpr.14200000000000003\linewidth-2\tabcolsep}}
\tbltoprule Dataset & Method & Accuracy  & Precision & Recall & F1-score & Cohen's kappa score\\
\tblmidrule 
\multicolumn{1}{p{\dimexpr(.14\linewidth-2\tabcolsep)}}{\multirow{7}{\linewidth}{\cAlignHack ESC-10}} &
  Baseline &
  0.843 (0.076)   &
  0.876 (0.061) &
  0.865 (0.056) &
  0.849 (0.066) &
  0.826 (0.084)\\
 &
  time\_stretch &
  0.948 (0.054) &
  0.958 (0.039) &
  0.96 (0.03) &
  0.952 (0.04) &
  0.942 (0.06)\\
 &
  pitch\_shift1 &
  0.951 (0.073) &
  0.966 (0.049) &
  0.971 (0.037) &
  0.96 (0.056) &
  0.945 (0.081)\\
 &
  pitch\_shift2 &
  0.95 (0.068) &
  0.961 (0.05) &
  0.965 (0.042) &
  0.957 (0.05) &
  0.944 (0.075)\\
 &
  drcomp &
  0.95 (0.068) &
  0.962 (0.047) &
  \textbf{0.971 (0.029)} &
  0.957 (0.05) &
  0.944 (0.075)\\
 &
  background &
  0.949 (0.065) &
  0.96 (0.049) &
  0.97 (0.03) &
  0.956 (0.05) &
  0.943 (0.073)\\
 &
  GAN &
  \textbf{0.965 (0.034)} &
  \textbf{0.975(0.024)} &
  0.965 (0.033) &
  \textbf{0.964 (0.034)} &
  \textbf{0.961(0.038)}\\\cline{1-1}\cline{2-2}\cline{3-3}\cline{4-4}\cline{5-5}\cline{6-6}\cline{7-7}
\multicolumn{1}{p{\dimexpr(.14\linewidth-2\tabcolsep)}}{\multirow{7}{\linewidth}{UrbanSound8K}} &
  Baseline &
  0.938 (0.025) &
  0.924 (0.041) &
  0.917 (0.036) &
  0.918 (0.038) &
  0.93 (0.029)\\
 &
  time\_stretch &
  0.954 (0.019) &
  0.944 (0.044) &
  0.941 (0.042) &
  0.941 (0.043) &
  0.947 (0.022)\\
 &
  pitch\_shift1 &
  0.927 (0.02) &
  0.905 (0.039) &
  0.889 (0.04) &
  0.893 (0.039) &
  0.917 (0.022)\\
 &
  pitch\_shift2 &
  0.95 (0.014) &
  0.931 (0.043) &
  0.931 (0.04) &
  0.929 (0.04) &
  0.943 (0.016)\\
 &
  drcomp &
  \textbf{0.995 (0.004)} &
  \textbf{0.996(0.003)} &
  \textbf{0.995 (0.004)} &
  \textbf{0.995 (0.004)} &
  \textbf{0.994(0.004)}\\
 &
  background &
  0.974 (0.024) &
  0.968 (0.033) &
  0.962 (0.031) &
  0.964 (0.032) &
  0.971 (0.027)\\
 &
  GAN &
  0.99 (0.005) &
  0.992 (0.003) &
  0.991 (0.004) &
  0.992 (0.004) &
  0.988 (0.005)\\\cline{1-1}\cline{2-2}\cline{3-3}\cline{4-4}\cline{5-5}\cline{6-6}\cline{7-7}
\multicolumn{1}{p{\dimexpr(.14\linewidth-2\tabcolsep)}}{\multirow{7}{\linewidth}{TUT}} &
  Baseline &
  0.687 (0.048) &
  0.682 (0.046) &
  0.671 (0.05) &
  0.669 (0.052) &
  0.648 (0.054)\\
 &
  time\_stretch &
  0.775 (0.047) &
  0.772 (0.046) &
  0.765 (0.048) &
  0.764 (0.047) &
  0.747 (0.053)\\
 &
  pitch\_shift1 &
  0.749 (0.031) &
  0.742 (0.032) &
  0.739 (0.03) &
  0.737 (0.032) &
  0.718 (0.034)\\
 &
  pitch\_shift2 &
  0.764 (0.027) &
  0.76 (0.028) &
  0.752 (0.032) &
  0.752 (0.03) &
  0.735 (0.03)\\
 &
  drcomp &
  \textbf{0.857 (0.02)} &
  \textbf{0.85 (0.029)} &
  \textbf{0.848 (0.031)} &
  \textbf{0.85 (0.029)} &
  \textbf{0.839(0.023)}\\
 &
  background &
  0.599 (0.093) &
  0.592 (0.093) &
  0.575 (0.09) &
  0.566 (0.095) &
  0.548 (0.105)\\
 &
  GAN &
  0.73 (0.05) &
  0.721 (0.05) &
  0.714 (0.045) &
  0.711 (0.05) &
  0.697 (0.053)\\
\tblbottomrule 
\end{tabulary}\par 
\begin{tablenotes}\footnotesize 
    
\item{Bold indicates the highest value of each metric}
\end{tablenotes}
\end{threeparttable}

\end{table*}

\subsection{Performance of the proposed method}InFigure~\ref{f-6b66d1fb3f34}, we present the confusion matrix obtained by training the suggested model on the augmented ESC-10 dataset.Figure~\ref{f-7f14e37d1ddb} shows the difference between the confusion matrices obtained by training the model on augmented and original ESC-10 dataset. In these matrices, the labels 'DO','RA','SE','BA','CL','PE','HE','CH','RO', and 'FI' represent the classes \textit{dog\_barking, rain, sea\_waves, baby\_cry, clock\_tick, person\_sneeze, helicopter, chainsaw, rooster and fire crackling }respectively. In Figure~\ref{f-7f14e37d1ddb}, off diagonal negative entries indicate that augmentation decreases the confusion between the concerned classes. On the other hand, the off diagonal positive values suggest that augmentation increases the confusion between concerned classes. For the main diagonal entries, the positive values indicate that augmentation improves the classification accuracy for the concerned class. The negative values suggest that augmentation deteriorates the classification accuracy for the concerned class.

\bgroup
\fixFloatSize{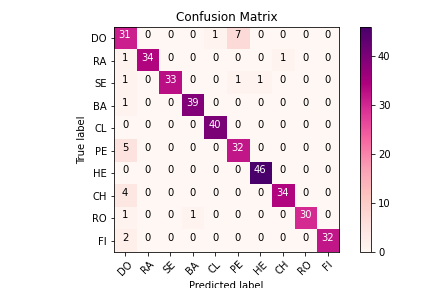}
\begin{figure}[!htbp]
\centering \makeatletter\IfFileExists{images/cm_with_gan.png}{\includegraphics{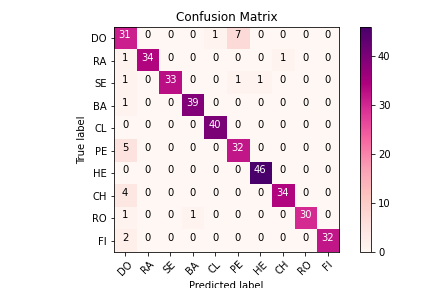}}{}
\makeatother 
\caption{{Confusion matrix with augmentation for ESC-10}}
\label{f-6b66d1fb3f34}
\end{figure}
\egroup

\bgroup
\fixFloatSize{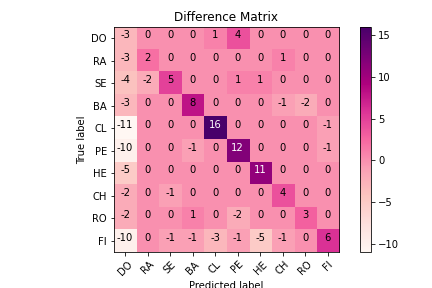}
\begin{figure}[!htbp]
\centering \makeatletter\IfFileExists{images/difference_matrix.png}{\includegraphics{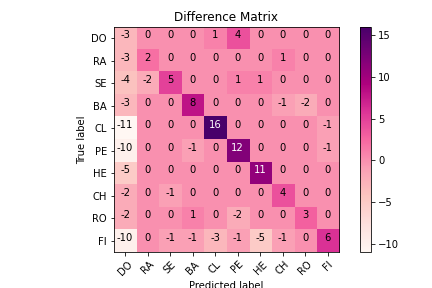}}{}
\makeatother 
\caption{{Difference matrix for ESC-10}}
\label{f-7f14e37d1ddb}
\end{figure}
\egroup
From the difference matrix it can be observed that augmentation improves the classification accuracy for all classes except \textit{dog\_barking}. In addition, it can be seen that augmentation has a deleterious result on the confusion between certain classes. For example, we notice that while augmentation reduces the confusion between sea\_waves and rain\textit{\space }classes, it increases the confusion between\textit{ sea\_waves }and\textit{ person\_sneeze }classes.

\subsection{Comparison with state-of-the-art}Table~\ref{tw-1ab8687c045a} shows the comparison of our model with state-of-the-art methods. It is evident from the table that our method outperforms state-of-the-art methods. The improvement in the performance of the proposed model can be attributed to the increased variance of the dataset introduced by the proposed augmentation using GAN. We can observe that GAN is superior to other augmentation methods in tackling imbalanced datasets like UrbanSound8K. However, the superior performance is not only due to the proposed augmentation, but also due to the high capacity CNN model with immense representational power employed in this work.
\begin{table}[!htbp]
\caption{{Comparison with state-of-the-art methods} }
\label{tw-1ab8687c045a}
\centering 
\begin{threeparttable}

\def\arraystretch{1}
\ignorespaces 
\centering 
\begin{tabulary}{\linewidth}{p{\dimexpr.52550000000000004\linewidth-2\tabcolsep}p{\dimexpr.1718\linewidth-2\tabcolsep}p{\dimexpr.1602\linewidth-2\tabcolsep}p{\dimexpr.14249999999999998\linewidth-2\tabcolsep}}
\tbltoprule \cAlignHack Model & \cAlignHack ESC-10 & \cAlignHack US8K & \cAlignHack TUT\\
\tblmidrule 
DCNN + regularization + augmentation\unskip~\cite{864172:20205986} &
  \cAlignHack 0.9494 &
  \cAlignHack 0.9537 &
  \cAlignHack -----\\
DCNN + attention\unskip~\cite{864172:20205998} &
  \cAlignHack 0.9575 &
  \cAlignHack 0.9752 &
  \cAlignHack -----\\
PiczakCNN\unskip~\cite{864172:20206010} &
  \cAlignHack 0.805 &
  \cAlignHack 0.727 &
  \cAlignHack -----\\
DCNN\unskip~\cite{864172:20206014} &
  \cAlignHack ----- &
  \cAlignHack 0.819 &
  \cAlignHack -----\\
SoundNet\unskip~\cite{864172:20205995} &
  \cAlignHack 0.921 &
  \cAlignHack ------ &
  \cAlignHack -----\\
Envnet-v2 + augmentation\unskip~\cite{864172:20205989} &
  \cAlignHack 0.917 &
  \cAlignHack 0.837 &
  \cAlignHack -----\\
DCNN + mixup + augmentation\unskip~\cite{864172:20206025} &
  \cAlignHack 0.917 &
  \cAlignHack 0.837 &
  \cAlignHack -----\\
CNN + attention + augmentation\unskip~\cite{864172:20205992} &
  \cAlignHack 0.932 &
  \cAlignHack ----- &
  \cAlignHack -----\\
CNN\unskip~\cite{864172:20205994} &
  \cAlignHack ----- &
  \cAlignHack ----- &
  \cAlignHack 0.597\\
GoogLeNet\unskip~\cite{864172:20206008} &
  \cAlignHack 0.91 &
  \cAlignHack 0.93 &
  \cAlignHack -----\\
Proposed Model + augmentation &
  \cAlignHack \textbf{0.965} &
  \cAlignHack \textbf{0.9897} &
  \cAlignHack \textbf{0.73}\\
\tblbottomrule 
\end{tabulary}\par 
\begin{tablenotes}\footnotesize 
    
\item{Bold indicates the highest value of the metric}
\end{tablenotes}
\end{threeparttable}

\end{table}

\section{Conclusion}
In this work, we developed the EnvGAN, the first application of GANs for adversarial generation of environmental sounds. Our experiments suggest that EnvGAN can generate sounds similar to the ones in three benchmark datasets - ESC-10, UrbanSound8K and TUT Urban Acoustic Scenes Development dataset. To evaluate the utility of EnvGAN in generating synthetic environmental sound, we compared the performance of a high capacity classifier on the original dataset and the augmented training set. The augmented training set in combination with the proposed high capacity deep learning model outstripped the state-of-the-art methods for ESC. In our experiments, we noticed that each augmentation set affects the model performance metrics for each class differently. This suggests that by using class conditional augmentation during training, the performance of the model could be further improved. That is, a validation set could be used to identify the augmentation which significantly improves the model's performance for each class and then the training data could be selectively augmented accordingly. This is an exciting avenue for further research.
\section*{Declaration of competing Interest}The authors declare that they have no known competing financial, general and institutional interests or personal relationships that could have appeared to influence the work reported in this paper.

\bibliographystyle{vancouver}

\bibliography{article}

\end{document}